\documentclass[aps, prl, showpacs, twocolumn, superscriptaddress]{revtex4}
\usepackage{graphicx}
\usepackage{amsmath}
\begin{document}
\title{Universality Classes in Constrained Crack Growth}
\date{\today}
\author{Knut S.\ \surname{Gjerden}}
\email{knut.skogstrand.gjerden@gmail.com}
\affiliation{Department of Physics,
Norwegian University of Science and Technology, N-7491 Trondheim,
Norway}
\author{Arne \surname{Stormo}}
\email{arne.stormo@gmail.com}
\affiliation{Department of Physics,
Norwegian University of Science and Technology, N-7491 Trondheim,
Norway}
\author{Alex \surname{Hansen}}
\email{alex.hansen@ntnu.no}
\affiliation{Department of Physics,
Norwegian University of Science and Technology, N-7491 Trondheim,
Norway}
\begin{abstract}
Based on an extension of the fiber bundle model we investigate numerically 
the motion of the crack front through a weak plane separating a 
soft and an infinitely stiff block.  We find that there are two regimes.
At large scales the motion is consistent with the pinned elastic line model
and we find a roughness exponent equal to $0.39\pm0.04$ characterizing it.
At smaller scales, coalescence of holes dominates the motion rather than 
pinning of the crack front.  We find a roughness exponent consistent with 
2/3, which is the gradient percolation value.  The length of the crack 
front is fractal on these smaller scales.  We determine its fractal 
dimension to be $1.77\pm0.02$, consistent with the hull of percolation 
clusters, 7/4. This suggests that the crack front is described by two 
universality classes: on large scales, the pinned elastic line one and on 
small scales, the percolation universality class.  
\end{abstract}
\maketitle

Schmittbuhl and M{\aa}l{\o}y \cite{sm97} were the first to study
experimentally the roughness of a crack front moving along a weak plane
in a material loaded under mode I conditions.  By sintering two sandblasted
plexiglass plates together and then plying them apart from one edge, they
were able to follow the motion of the crack front moving along the 
sintered boundary between the two plates.  The rough crack front turned
out to be self affine, i.e., its height-height correlation function 
$\langle (h(x+\Delta x)-h(x))^2\rangle$ scaled as $|\Delta x|^\zeta$
where $\zeta$ is the rougness exponent.  $h(x)$ is the position of the crack
front with respect to a base line orthogonal to the average crack growth
direction and $x$ is the coordinate along this base line.  The roughness
exponent was found to be $\zeta=0.55\pm0.05$.  A couple of years before this
work, Schmittbuhl et al.\ \cite{srvm95} studied numerically a model of
such constrained crack growth based on regarding the motion of the crack
front as that of a pinned elastic line, the fluctuating line model. This
work was based on an earlier idea by Bouchaud et al.\ \cite{bblp93}.
The conclusion of Schmittbuhl et al.\ 
was that the front should be rough and self affine,
characterized by a roughness exponent $\zeta=0.35\pm0.05$.  This
value was later refined to $0.388\pm0.002$ by Rosso and Krauth \cite{rk02}.  
The large discrepancy between the numerical and experimental results ---
the latter having been refined to $\zeta=0.63\pm0.03$ by improving the 
statistical analysis \cite{dsm99} --- spurred
a lively quest for an explanation that only today seems to converge towards
a satisfactory understanding of the underlying physics , see e.g.\
\cite{b09,bb11} for a review. 

An attempt at finding an explanation for the large roughness exponent
seen in the experiments that was not based on some variation on the 
fluctuating line model was forwarded by Schmittbuhl et al.\ \cite{shb03}. 
The underlying idea here was that the crack front did not advance due to
a competition between effective elastic forces and pinning forces at the 
front, but by {\it coalescence\/} of damage in front of the crack with the
advancing crack itself. Such an idea had been put forwards in a more
general context by Bouchaud et al.\ \cite{bbfrr02} the year before. 
By using a grid of linearly elastic fibers 
connected to a soft elastic half space, Schmittbuhl et al.\ 
found a rougness exponent
of $\zeta=0.60\pm0.05$, which was consistent with the experimental results. 
The crack advance mechanism in this model was analyzed in terms of 
a correlated gradient percolation process where coalescence is the 
dominating mechanism.

Recently, Santucci et al.\ \cite{sgtshm10} reanalyzed data from a number
of earlier studies, including \cite{dsm99}, finding that the crack
front has {\it two scaling regimes:\/} one small-scale regime described
by a roughness exponent $\zeta_-=0.60\pm0.05$ and a large-scale 
regime described by a roughness exponent $\zeta_+=0.35\pm0.05$.  

We suggest in this letter that there are indeed two competing mechanisms
involved in generating the scaling properties seen in the roughness of
the crack front: on small scales coalescence dominates, whereas on large
scales, the fluctuating line picture is correct.  There is a crossover
between the regimes where either of the two mechanisms dominate 
associated with a well-defined crossover length scale.  

Our numerical work strongly suggest that the coalescence mechanism 
seen at small scales is controlled by the ordinary percolation critical
point, leading to percolation exponents.  This is in constrast to 
Schmittbuhl et al.\ \cite{shb03} who suggested a new percolation-like 
universality class, but consistent with recent theoretical
work on fracture in the fuse model \cite{szs12}.  

We note an earlier attempt at explaining the two scaling regimes seen in 
Ref.\ \cite{sgtshm10} based solely on the fluctuating line model
\cite{lsz10}.  Here Laurson et al.\ \cite{lsz10} relate the crossover to
the Larkin length scale of the crack front \cite{lo79}.  This is in this
context the length scale at which the roughness of the front is comparable 
to the correlation length inherent to the pinning disorder. 

Our numerical system is a fiber bundle model \cite{phc10} 
closely related to the one used in \cite{shb03}
and introduced by Batrouni et al.\ the year before \cite{bhs02}. 
$L\times L$ elastic fibers are placed in a square lattice between two
clamps.  One of the clamps is infinely stiff whereas the other has a finite
Young modulus $E$ and a Poisson ratio $\nu$.  All fibers are equally long
and have the same elastic constant $k$.  We measure the position of the stiff
clamp with respect its position when all fibers carry zero force, $D$.  The
force carried by the fiber at position $(i,j)$, where $i$ and $j$ are
coordinates in a cartesian coordinate system oriented along the 
edges of the system, is then
\begin{equation}
\label{base}
f_{(i,j)}=-k(u_{(i,j)}-D)\;,
\end{equation}
where $u_{(i,j)}$ is the fiber's elongation.  
The fibers redistribute the forces
they carry through the response of the clamp with finite elastisity. 
The redistribution of forces is accomplished by using the Green function
connecting the force $f_{(m,n)}$ acting on the clamp from fiber 
$(m,n)$ with the
deformation $u_{(i,j)}$ at fiber $(i,j)$, \cite{j85}
\begin{subequations}
\label{green}
\begin{align}
u_{(i,j)}&= \sum_{(m,n)}G_{(i,j),(m,n)}f_{(m,n)},\\
G_{(i,j),(m,n)}& = \nonumber \\
\frac{1-\nu^2}{\pi E a^2}&\int^{a/2}_{-a/2} dx\ \int^{a/2}_{-a/2} dy\
\frac{1}{|\vec{r}_{(i,j)}-\vec{r}_{(m+x,n+y)}|}. \label{greenb}
\end{align}
\end{subequations}
where $a$ is the distance between neighboring fibers. 
$\vec{r}_{(i,j)}-\vec{r}_{(m,n)}$ is the distance between fibers $(i,j)$ 
and $(m,n)$ and the integration runs over the $a\times a$ square around 
fiber $(m,n)$.  This equation set is solved using a Fourier accelerated 
conjugate gradient method \cite{bhn86,bh88}. 

\begin{figure*}[ht]
\includegraphics{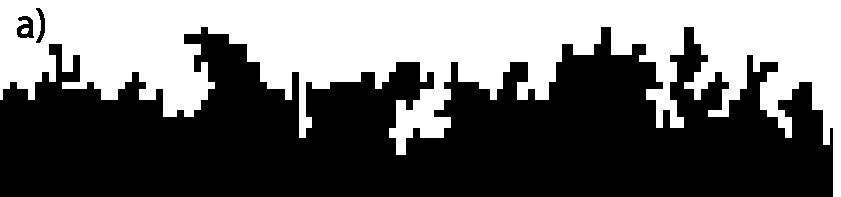} 
\includegraphics{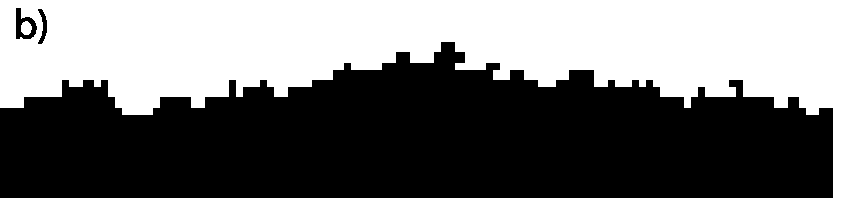}
\caption{Parts of the crack fronts obtained in 
the simulations. The color black represents broken 
bonds and white represents unbroken bonds. 
To the left, a) is an example from a system with 
a high rescaled Young modulus $e$ being driven forwards 
(upwards) primarily by coalescence with damage forming 
ahead of the crack front. (b) is an example from a system 
with a low $e$. This system is being 
driven forward by damage forming on the crack front.
Small $e=2\times10^{-3}$ is equivalent to a larger system than one with a
large $e=0.8$.}
\label{fig1}
\end{figure*}

We note that the Green function, Eq.\ (\ref{greenb}), is proportional to
$(Ea)^{-1}$.  The elastic constant of the fibers, $k$, must be 
proportional to $a^2$.  The linear size of the system is $aL$.  Hence,
by changing the linear size of the system without changing the discretization
$a$, we change $L \to \lambda L$ but leave $(Ea)$ and $k$ unchanged.  
If we on the other hand change the discretization without changing the 
linear size of the system, we simultaneously set $L\to\lambda L$,
$(Ea)\to\lambda (Ea)$ and $k\to k/\lambda^2$. We define the scaled
Young modulus $e=(Ea)/L$.  Hence, changing $e$ without changing $k$ is
equivalent to changing $L$ --- and hence the linear size of the system ---
while keeping the elastic properties of the system constant \cite{sgh12}.

Bonds are broken by using the quasistatic approach \cite{h05}.  That is,
we assign to each fiber $(i,j)$ a threshold value $t_{(i,j)}$.  
Bonds are broken one
by one by each time identifying 
$\max_{(i,j)}(f_{(i,j)}/t_{(i,j)})$ for $D=1$. This ratio
is then used to read off the value $D$ at which the next fiber breaks.

In the constrained crack growth experiments of Schmittbuhl and M{\aa}l{\o}y
\cite{sm97}, the two sintered plexiglass plates were plied apart from one
edge.  In the numerical modeling of Schmittbuhl et al.\ \cite{shb03},
an asymmetric loading was accomplished by introducing a linear 
gradient in $D$. Rather than implementing an asymmetric loading, we
introduce a gradient in the threshold distribution,
$t_{(i,j)}=g j + r_{(i,j)}$, where $g$ is the gradient and $r_{(i,j)}$
is a random number drawn from a flat distribution on the unit interval.
In the limit of large Young modulus $E$, this system becomes equivalent to
the gradient percolation problem \cite{srg85,gr05}.

In order to follow the crack front as the breakdown process develops, we
implement the ``conveyor belt" technique \cite{drp01,gsh12} where a new upper
row of intact fibers is added and a lower row of broken fibers removed
from the system at regular intervals.  This makes it possible to follow
the advancing crack front indefinitely.  The implementation has been
described in detail in \cite{gsh12}.  

\begin{figure}[ht]
\includegraphics[width=3.5in]{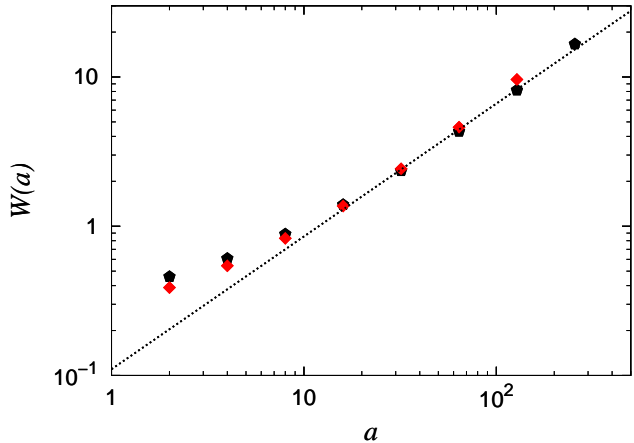}
\caption{Averaged wavelet coefficient $W(a)$ based on transforming 
$h_i$ vs.\ $a$ for 
$e=7.8125\times 10^{-4}$, $L=256$ ($g=0.00625$, averaged over 4200 fronts) 
and $L=512$ ($g=0.003125$, averaged over 850 fronts). The slope of the 
straight line is $0.39+1/2$.}
\label{fig2}
\end{figure}

Figure \ref{fig1} shows two examples of typical crack
fronts representative of a stiff (high $e=0.8$) and a soft system 
(low $e=2\times10^{-3}$).  In both cases, the front is viewed from above 
(and propagation corresponds to an upwards 
movement in the figure). The two fronts are quite different.  Even though
we change the Young modulus $E$ between the two crack fronts, this is
equivalent to changing the scale at which the fronts are viewed: Small
$e$ corresponds to a large scale and {\it vice versa.\/}

We identify the crack front by first eliminating all islands of
surviving fibers behind it and all islands of failed fibers in front
of it.  We measure ``time" $n$ in terms of the number of failed fibers.  
After an initial period, the system settles into a steady state. We then
record the position of the crack front $j=j(i, n = 0)$ after having set 
$n=0$.  We then define the position at later times $n>0$ relative to this
initial position,
\begin{equation}
\label{defh}
h_i(n)=j(i,n)-j(i,n=0)\;.
\end{equation}
This is the same definition as was used by Schmittbuhl and M{\aa}l{\o}y
\cite{sm97}. The front as it has now been defined will contain overhangs.
That is, there may be multiple values of $h_i(n)$ for the same $i$ and $n$
values.  We only keep the largest $h_i(n)$, i.e.\ we implement the Solid-on-Solid
(SOS) front. 

We define the average position of the front as $\langle h(n)\rangle=
\sum_{i=1}^L h_i(n)/L$ and the front width as 
$w(n)^2=\sum_{i=1}^L(h_i(n)-\langle h(n)\rangle)^2/L$. 

We will in the following explore the model going from large to small
values of the Young modulus $E$ while keeping $k$ fixed.  As already
discussed, this is equivalent to changing the system size $L$ while keeping
$E$ fixed.

\begin{figure}[ht]
\includegraphics[width=3.5in]{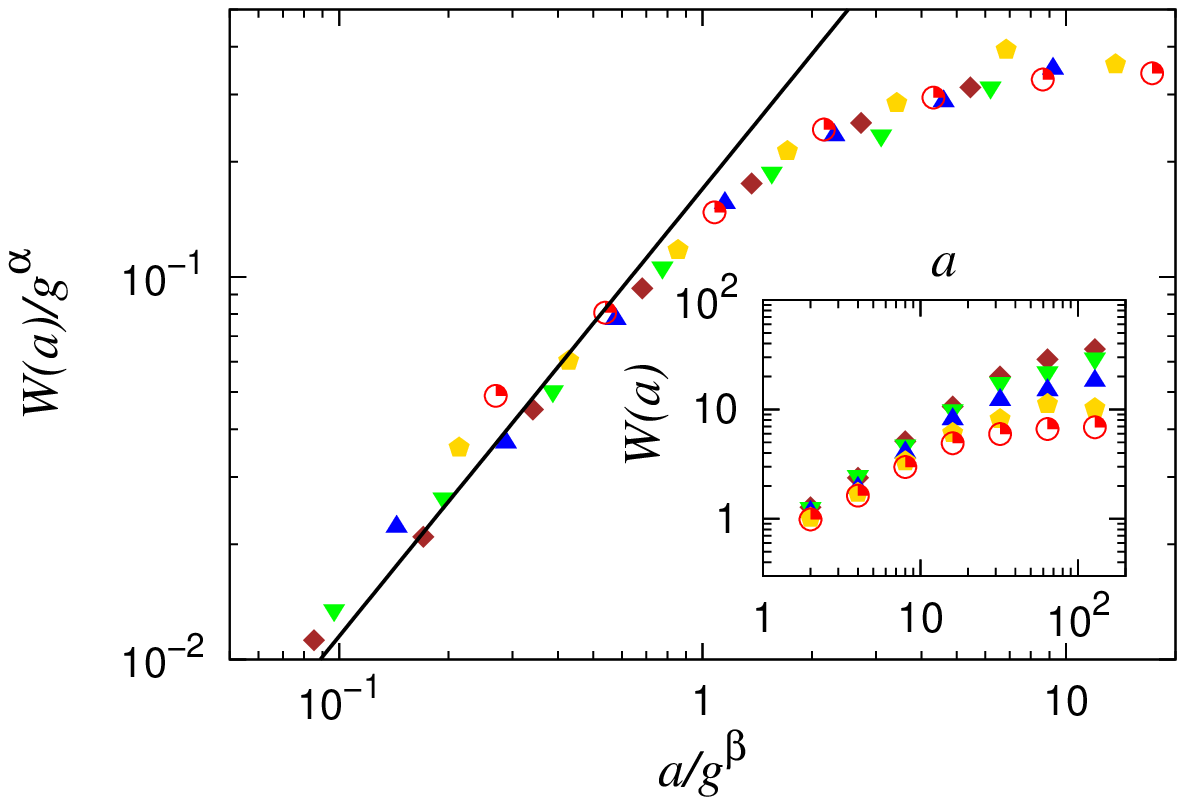}
\includegraphics[width=3.5in]{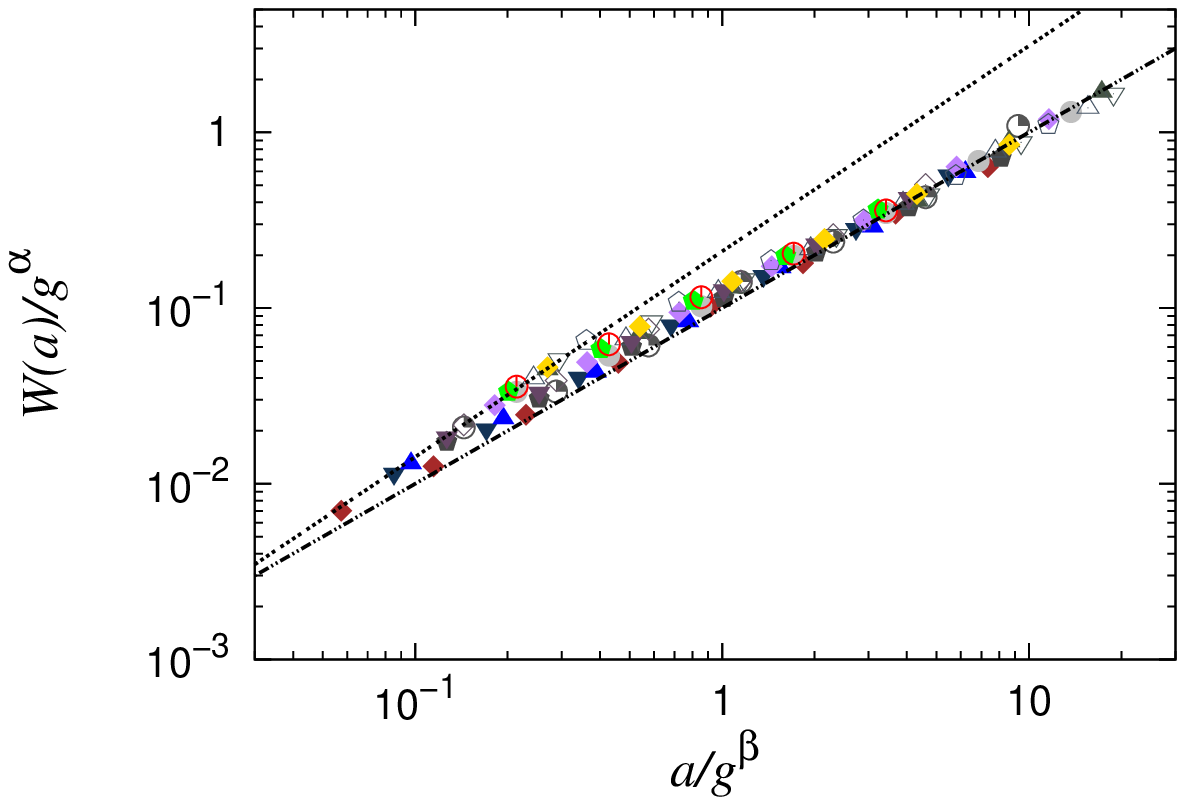}
\caption{a) Averaged wavelet coefficients $W(a)$ of the SOS front $h_i$
vs.\ scale $a$. The straight line has slope $2/3+1/2=6/7$.  The
insert shows the unscaled data. The data are
based on $e=3.125$, $L=256$  ($g=0.004$, 0.005, 0.01, 0.02 and 0.03).
b) Average wavelet coefficients $W(a)$ of the
SOS fronts $h^0_i$ where the overhangs have been removed.  
The long-dashed line has slope $2/3+1/2=6/7$ and the short-dashed 
line has slope $1/2+1/2=1$.  In both a) and b), $\alpha=\beta=4/7$.
The data are based on  $e=3.125$, $L=64$ ($g=0.018$ and 0.02, averaged
over 1426 fronts),
$L=128$ ($g=0.008$, 0.01, 0.03, and 0.05, averaged over 100 to 500 fronts), 
$L=256$ ($g=0.004$, 0.005, 0.008, 0.01, 0.015, 0.02, 0.025, 0.03, and 0.035,
averaged over 30 to 500 fronts), and $L=512$ 
($g=0.002$, averaged over 50 fronts).}
\label{fig3}
\end{figure}

We analyze the fracture fronts in the following using the average wavelet 
coefficient (AWC) method \cite{mrs97,shn98}.  
We transform the front $h_i(n)$ using the Daubechies-4 wavelet,
$h_i(n)\to w_{a,b}$ where $a$ is scale and $b$ is position (and we have
suppressed the $n$ dependency).  We then
average $|w_{a,b}|$ for each length scale $a$ over position $b$, and
if $h_i(n)$ is self affine with a roughness exponent $\zeta$, we have
\begin{equation}
\label{waveletave}
W(a) =\langle | w_{a,b}|\rangle_b\sim a^{\zeta+1/2}\;.
\end{equation}

We start by considering systems with small scaled Young modulus $e$.  This is
a soft system at small length scales --- or equivalently, a stiffer system at 
large length scales. We set $e=7.8125\times 10^{-4}$.  The fronts in this
regime have an appearance as in Fig.\ \ref{fig1}b. 
In Fig.\ \ref{fig2}, we plot the averaged wavelet 
coefficient $W(a)$ against the scale $a$.  The data follow a power law,
$W(a) \sim a^{0.39+1/2}$, leading to a roughness exponent of
\begin{equation}
\label{softzeta}
\zeta_+ = 0.39\pm 0.04\;,
\end{equation}
entirely consistent with the large scale rougness exponent measured by
Santucci et al.\ \cite{sgtshm10}, $\zeta_+=0.35\pm0.05$.

We now turn to large scaled Young modulus, i.e.\ stiff systems --- 
or equivalently, softer systems on small length scales. Hence, fronts
appear as in Fig.\ \ref{fig1}a.  We set in the 
following $e=3.125$.  The corresponding plot of averaged wavelet coefficient
$W(a)$ vs.\ scale $a$ is shown in the insert in Fig.\ \ref{fig3}a.  The
different curves correspond to different gradients $g$ and $L$.
Repeating the analysis of Sapoval et al.\ \cite{srg85,hbrs07} for gradient 
percolation, 
we assume that the front has an isotropic correlation 
length $\xi$ associated with it. In the case of gradient percolation, this 
correlation length is related to the gradient through the relation 
$\xi \sim g^{-\nu/(1+\nu)}=g^{-4/7}$, where $\nu=4/3$ is the percolation 
correlation length exponent.  As was argued in Ref.\ \cite{shb03}, the same 
analysis may be repeated for correlated systems such 
as the present one, but with a 
possibly different $\nu$.  We would then expect data collapse in Fig.\ 
\ref{fig3}a by rescaling $W(a)\to W(a)/\xi = W(a)/g^{-\alpha}$ and $a\to
a/\xi=a/g^{-\beta}$, and where $\alpha=\beta=\nu/(1+\nu)$.  We show in 
Fig.\ \ref{fig3}a, the data collapse that ensues when choosing
$\alpha=\beta=4/7$, i.e., the ordinary {\it uncorrelated percolation\/} value
$\nu/(1+\nu)$, where $\nu=4/3$.
There are two distinct regions in the figure.  For large values of 
$a/g^{-\beta}$, $W(a)/g^{-\alpha}$ is independent of $a/g^{-\beta}$.  Hence,
the front has the character of uncorrelated noise. We will discuss this further
on.  For small $a/g^{-\beta}$,
the data follow a power law.  We show in the figure (Fig.\ \ref{fig3}a)
a straight line with slope
$7/6=2/3+1/2$.  Hence, the data are consistent with the fronts being self
affine on these scales with roughness exponent 
\begin{equation}
\label{stiffzeta}
\zeta_-=2/3\;,
\end{equation} 
a value 
consistent with gradient percolation \cite{hbrs07} --- and consistent with 
the experimental value reported by Santucci et al.\ \cite{sgtshm10}, 
$\zeta_-=0.60\pm0.05$.

In order to further the analysis, we follow the procedure in \cite{hbrs07}
by smoothening the fronts by removing the jumps due to the overhangs through 
the transformation
\begin{eqnarray}
\label{trans}
h_i(n)\to h_i^k(n) =\nonumber\\ 
\sum^i_{m=0} 
\textrm{sgn}[h_{m+1}(n)-h_m(n)]\ |h_{m+1}(n)-h_m(n)|^k\;,\nonumber\\
\end{eqnarray}
when $k\to 0$ \cite{bh07}.  Hence, all steps are equal to one in $h_i^0(n)$,
all overhangs have been removed.  We ensure by this procedure that the 
scaling properties of the roughness are due to self affinity and {\it not\/}
due to the overhangs which are prevalent in this regime.  

We plot in Fig.\ \ref{fig3}b the rescaled $W(a)/g^{-\alpha}$ based on 
transforming $h^0_i$ vs.\ the rescaled $a/g^{-\beta}$.  We have set
$\alpha=\beta=4/7$, consistent with ordinary gradient percolation. 
The two straight lines that have been added to the figure have slopes 
$2/3+1/2=7/6$ and $1/2+1/2=1$ respectively.  On small scales, the fronts
are then self affine with a roughness exponent $\zeta_-=2/3$ --- the gradient
percolation value \cite{hbrs07}.  
On larger scales and with the overhangs removed, one would naively have expected
to observe the fluctuating line regime characterized 
by a roughness exponent $\zeta_+=0.39$.  However, by removing the overhangs,
the effective roughness exponent one measures is $\max(1/2,\zeta_+)$,
which in this case is 1/2 \cite{bh07}.
Fig.\ \ref{fig3}b, then, shows the crossover from a roughness exponent
consistent with ordinary gradient percolation to a plain random walk exponent
which is a result of the smoothening process.

Hence, we conclude that the scaling properties seen on small scales are
consistent with uncorrelated gradient percolation.  The analysis of 
Schmittbuhl et al.\ \cite{shb03} suggested a correlated gradient
percolation process.  This is one step further.  We suggest that the 
process is an uncorrelated gradient percolation process on small scales
where coalescence is the dominating process.

\begin{figure}[ht]
\includegraphics[width=3.5in]{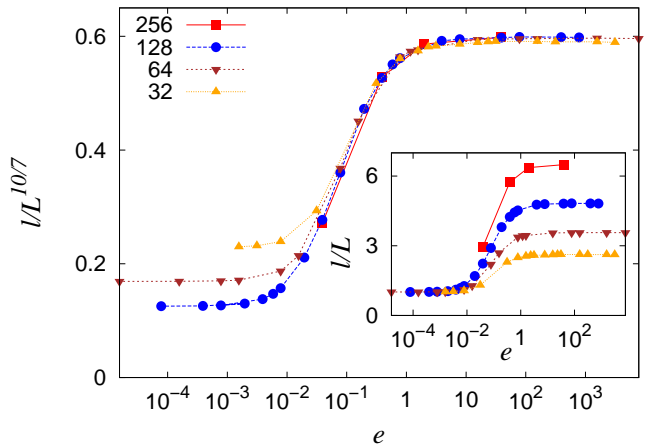}
\caption{Length of the fracture front $l$ scaled by system size 
plotted against effective Young modulus $e$.  The main figure 
shows data collapse for $l/L^{10/7}$ for large $e$ values and the
insert shows data collapse for small $e$ values for $l/L^1$.  We set
$g=1.6/L$. The data are based on 425000 fronts for $L=32$, 286482 fronts 
for $L=64$, 282022 fronts for $L=128$ and 32117 fronts for $L=256$.}
\label{fig4}
\end{figure}

Roughness exponents are notoriously difficult to measure.  The data
presented in Fig.\ \ref{fig3} are not of sufficient quality to warrant
firm conclusions on the small-scale universality class by themselves.  We
therefore in the following measure the {\it fractal dimension\/} of the
fronts.  Leaving the SOS assumption, we now follow the front as shown in
Fig.\ \ref{fig1}.  The front has a length $l$. For small values of the
scaled Young modulus $e$, there are no (or very few) overhangs and we
expect $l$ to be proportional to $L$: it is not fractal.  However, for large
$e$ where overhangs are prevalent, we do expect it to be fractal.  We assume
that there is a correlation length $\xi$ and that the front is fractal
up to this scale.  Hence, the length of the front $l$ then scales as
\begin{equation}
\label{fractal1}
l\sim \xi^{D_f}\ \left(\frac{L}{\xi}\right)\;.
\end{equation}
From Sapoval et al.\ \cite{srg85}, we know that $\xi\sim g^{-\nu/(1+\nu)}$. We
now set $g=c/L$, where $c$ is a constant.  Hence, we find
\begin{equation}
\label{fractal2}
l\sim L ^{(\nu D_f+1)/(\nu+1)}\;.
\end{equation}
If we are dealing with ordinary gradient percolation, $\nu=4/3$ and $D_f=7/4$
\cite{gr05}.  Hence, we expect $(\nu D_f+1)/(\nu+1)=10/7\approx 1.43$.  We
show in Fig.\ \ref{fig4}, $l/L^{10/7}$ as a function of the scaled Young
modulus, $e$.  For large values of $e$, there is excellent data collapse.  
For small values of $e$, there is data collapse when $l/L$ is plotted against
$e$, see the insert of Fig.\ \ref{fig4}, indicating that the front is not
fractal in this regime. We are seeing the crossover from the 
coalescence-dominated regime to the fluctuating line regime as $e$ moves from
large to smaller values.  Since $e=E/L$, we may keep $E$ fixed and change $e$
by changing $L$.  Hence, the coalescence regime is a small-scale regime whereas
the fluctating line regime is a large-scale regime. 

If we work backwards and search for the best scaling exponent to produce data
collapse in Fig.\ \ref{fig4}, we find $l/L^{1.44}$.  This gives 
$D_f=1.77\pm0.02$, entirely consistent with the percolation value $7/4$.

Hence, we have through the use of a single model based on the fiber
bundle model with a soft clamp and a gradient in the breaking threshold,
been able to identify two mechanisms by which the fracture front propagates.
On small scales, it is coalescence of damage the dominates, whereas on large
scales the front advances as descibed by the fluctuating line model.  The 
coalescence regime is in the universality class of ordinary percolation.

We are grateful for discussions and suggestions by D.\ Bonamy, K.\
J.\ M{\aa}l{\o}y and K.\ T.\ Tallakstad.  
We thank the Norwegian Research Council for financial 
support.  Part of this work made use of the facilities of HECToR, the UK's 
national high performance computing service, which is 
provided by UoE HPCx Ltd at the University of 
Edinburgh, Cray Inc and NAG Ltd, and funded by 
the Office of Science and Technology through 
EPSRC's High End Computing Programme. 


\end{document}